\documentclass[graybox,natbib,numbers,envcountsect]{svmult}

\usepackage{type1cm}          
                              
\usepackage{makeidx}          
\usepackage{graphicx}         
                              
\usepackage{multicol}         
\usepackage[bottom]{footmisc} 

\usepackage{newtxtext}
\usepackage[varvw]{newtxmath} 

\def\sepdot{$\cdot$ }
\def\eqref#1{\mbox{\rm(\ref{#1})}}
\newcommand{\td}{\,\mathrm{d}}

\begin{document}

\title*{Effect of physical aging on the flexural creep in 3D printed thermoplastic}

\author{\bf Marcel Fischbach and Kerstin Weinberg}
\authorrunning{M. Fischbach and K. Weinberg}

\institute{Marcel Fischbach \sepdot Kerstin Weinberg \at Chair of Solid Mechanics, {University of Siegen, Paul-Bonatz-Straße 9-11, 57076 Siegen, Germany} \email{marcel.fischbach@uni-siegen.de, kerstin.weinberg@uni-siegen.de}}

\maketitle

\abstract{Extrusion-based 3D printing has become one of the most common additive manufacturing methods and is widely used in engineering. This contribution presents the results of flexural creep experiments on 3D printed PLA specimens, focusing on changes in creep behavior due to physical aging. It is shown experimentally that the creep curves obtained on aged specimens are shifted to each other on the logarithmic time scale in a way that the theory of physical aging can explain. The reason for the physical aging of 3D printed thermoplastics is assumed to be the special heat treatment that the polymer undergoes during extrusion. Additionally, results of a long-term flexural creep experiment are shown, demonstrating that non-negligible creep over long periods can be observed even at temperatures well below the glass transition temperature. Such creep effects should be considered for designing components made of 3D printed thermoplastics.}

\section{Introduction}\label{sec:Intro}
Over the past decade, 3D printing technologies have become essential to modern manufacturing processes. Previously used primarily for rapid prototyping, 3D printing technologies are now employed to produce end-use parts in various applications, particularly for complex geometries \cite{Tan2020}. Since it is often necessary to prove that these printed parts can withstand the prevailing loads, knowledge of their short- and long-term material properties is required.

\begin{figure}[tb]
  \sidecaption
  \includegraphics[scale=1]{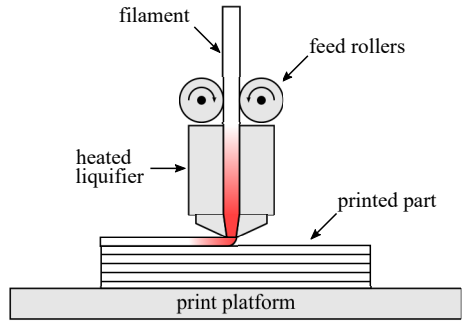}
  \caption{Schematic of extrusion-based 3D printing}
  \label{fig:Extruder}
\end{figure}

One of the most widely used 3D printing processes is fused filament deposition (FDM), in which plastic parts are built up layer by layer from a thermoplastic polymer filament. To generate the geometry of a single layer, the filament is heated above its melting temperature in an extruder and deposited through a nozzle on a build plate or a previously printed layer, Fig.~\ref{fig:Extruder}. When the term 3D printing is used in the following, we refer to this extrusion-based process. The most commonly used filament materials for this type of 3D printing are polylactide and acrylonitrile butadiene styrene \cite{Tan2020}, abbreviated hereafter as PLA and ABS, respectively.

Depending on the chosen print settings, e.g.~nozzle and platform temperature, print speed, infill density and infill pattern, to name only a few, printed parts have specific mechanical properties. Numerous studies have been conducted to quantify the relationship between chosen print parameters and the properties of 3D printed parts. For example, Ahn~\etal~\cite{Ahn2002} investigated the effect of selected print parameters on the tensile strength of 3D printed ABS using a two-level experimental design and analyzed the effect of print raster orientation on the tensile and compressive strength of printed compared to injection molded test specimens. Wittbrodt and Pearce~\cite{Wittbrodt2015} investigated the correlations between PLA filament color, nozzle temperature, degree of crystallinity, and yield strength of 3D printed tensile specimens. Fernandez-Vicente~\etal~\cite{Fernandez-Vicente2016} and Rismalia~\etal~\cite{Rismalia2019} analyzed the influence of common infill patterns and infill density on the tensile strength and tensile modulus of 3D printed specimens made of ABS and PLA, respectively. Reppel and Weinberg~\cite{Reppel2018} investigated the qualitative rupture behavior of printed tensile specimens made of thermoplastic polyurethane and modeled their deformation behavior using hyperelastic material models. Akhoundi~\etal~\cite{Akhoundi2020} studied the effects of nozzle temperature and heat treatment (annealing) on the crystallinity, the interlayer and intralayer adhesion, cf.~\cite{Tan2020}, and the mechanical properties of 3D printed tensile specimens of high-temperature PLA (HTPLA). Khosravani~\etal~\cite{Khosravani2022} analyzed the stress-strain behavior, the elastic modulus and tensile strength of 3D printed PLA specimens as a function of print raster orientation and two different print speeds.

In the field of time-dependent, i.e.~dynamic, viscoelastic and viscoplastic properties of printed thermoplastics, numerous contributions can also be found in literature, e.g.~\cite{Zhang2018,Mohammadizadeh2018,Tezel2019,Aghayan2021,Ye2021}. However, to our knowledge, there are none on the influence of process-induced physical aging on the creep behavior of 3D printed thermoplastics under usual production conditions.

Physical aging is a specific type of polymer aging that occurs in both amorphous and semi-crystalline thermoplastics, which are rapidly cooled below their glass transition temperature \cite{White2006}. In the resulting non-equilibrium state of low but not vanishing molecular mobility, the polymer gradually evolves toward thermodynamic equilibrium. As a result, this slow, time-dependent, and asymptotic process leads to a measurable stiffening of the polymer, which is particularly noticeable in the creep behavior, cf.~\cite{StruikPhD1977,Hastie1991,Sullivan1993,Pierik2020,Hodge1995,Hermida2018}. Additionally, the polymer's density and yield stress increase during aging while its impact strength decreases \cite{Hastie1991}.

As detailed in \cite{StruikPhD1977}, rapid cooling of polymers below their glass transition temperature can be artificially induced to study the effects of physical aging in a controlled manner. However, in addition to this academic approach, Struik~\cite{StruikPhD1977} argued that similar cooling conditions could also prevail during the processing of plastics by extrusion or injection molding. Struik demonstrated from creep experiments on injection molded PVC samples that those can initiate significant physical aging. Thus, concerning the special heat treatment to which the polymer filament is subjected during extrusion in 3D printing, the question arises as to whether and to what extent 3D printed components exhibit physical aging. Therefore, we address here the question of physical aging experimentally. Using a three-point bending setup, we investigate the flexural creep properties of 3D printed PLA specimens.

In this contribution, we briefly introduce the phenomena of physical aging and its relation to the viscoelasticity of thermoplastic polymers in Sect.~\ref{sec:Theory}. Then, in Sect.~\ref{sec:MandM}, the preparation of the specimens and the conducted experiments are explained. After the presentation of the experimental results in Sect.~\ref{sec:Results}, we discuss the measured results and critically evaluate the experimental procedure in Sect.~\ref{sec:Disc}.

\section{Theoretical Background}\label{sec:Theory}
In this section, the essence of physical aging and its linkage to the creep of thermoplastics is presented. In plastics, the term \textit{creep} refers to a time- and temperature-dependent increase in deformation under constant loading which is typically described by the viscoelastic compliance, i.e. the reciprocal of the material's stiffness.

\subsection{Viscoelasticity of Thermoplastics}\label{sec:Visco}
Viscoelasticity describes the time-dependent mechanical properties of polymers at temperatures below melting temperature. The macroscopic material behavior is typically symbolized by a general Maxwell model, i.e. a number of $N$ spring-damper elements in parallel to an elastic spring. For linear elastic and viscous relations and for $N \to \infty$, the dependence of the material's compliance $J$ from time $t \in [0,\infty)$ is given by the creep curve
\begin{align}\label{eqn:MaxwellCreepCurve}
  J(t) = J_0 + \int_0^\infty f(\tau) \left( 1-\exp\left(-\frac{t}{\tau}\right) \right) \td \tau \,.
\end{align}
Here $f(\tau)$ denotes the (measured or modeled) retardation spectrum and $J_0$ is the initial compliance.

Experimentally, such creep curves are obtained by measuring the history of strain $\varepsilon(t)$ related to the loading with stress $\sigma(t)$,
\begin{align}\label{eqn:DefCompliance}
   J(t) = \frac{\varepsilon(t)}{\sigma(t)} \,.
\end{align}
Because of the exponential relation \eqref{eqn:MaxwellCreepCurve}, the creep curve is typically plotted over a logarithmic time axis, $\log(t)$. Since the degree of molecular mobility essentially determines the viscosity of microscopic polymeric chain networks, the macroscopic behavior of a polymer is characterized by a strong temperature dependence. For thermo-rheologically simple materials, however, it is assumed that temperature only affects the velocity of molecular movements but not their type and number. Therefore, the shape of experimentally obtained creep curves remains the same at different temperatures and only their position on the time scale is different.

\begin{figure}[b]
  \centering
  \includegraphics[scale=1.0]{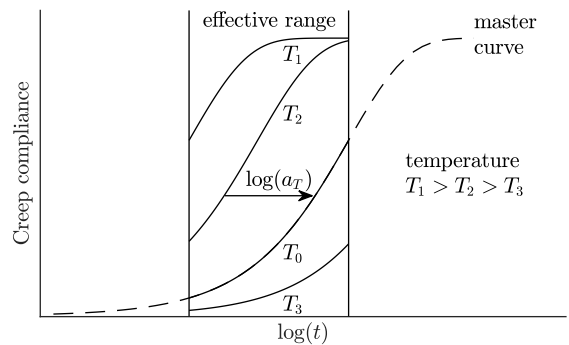}
  \caption{Master curve generation by means of time-temperature superposition (TTSP)}
  \label{fig:TTSP}
\end{figure}

This implies that from creep curves measured at different temperatures but within the same time interval a master curve at a reference temperature $T_0$ can be obtained by applying a time shift $\log(a_T)$ as illustrated in Fig.~\ref{fig:TTSP}. Depending on the reference temperature, the shift factor $a_T$ can be expressed as a function of temperature $T$ by the well-known Williams-Landel-Ferry (WLF) equation or an Arrhenius approach~\cite{Grellmann2013}. This method, known as \textit{time-temperature superposition principle} (TTSP) is commonly used to determine the rapid and long-term creep behavior of viscoelastic solids. Instead of conducting isothermal creep tests over inconvenient time periods, short-term tests are carried out at different temperatures and their results are combined. We remark that for thermo-rheologically complex materials a TTSP master curve does not result from horizontal shifting alone. Here,  additional vertical shifts may be required \cite{Pierik2020}.

The common rheological models with spring-damper arrangements can cover the full range of the material's viscoelastic behavior according to Eq.~\eqref{eqn:MaxwellCreepCurve}. A simplification, which corresponds to a linearization over short logarithmic time spans, gives a model which was proposed as early as 1863 by Kohlrausch~\cite{Kohlrausch1863}
\begin{equation}\label{eqn:KWW}
  J(t) = J_0\,\exp\left(\left(\frac{t}{\tau}\right)^m\right) \;.
\end{equation}
Here, $J_0$ describes the instantaneous creep response again, whereas $\tau$ is a characteristic retardation time associated with the active creep mechanism. This decay time increases with isothermal aging, which leads to a stretching of the curve. Exponent $m$ is the Kohlrausch coefficient, $0 < m\leq 1$, which is associated with the dispersion of retardation times. It is about $1/3$ for polymers \cite{StruikPhD1977}.

For obvious reasons, Eq.~\eqref{eqn:KWW} cannot represent the long-term creep behavior of glassy polymers over the entire retardation spectrum. Nonetheless, it is well suited for the representation of short-time creep and, in particular, for the practical determination of horizontal and vertical shifts of experimentally obtained creep curves. Therefore, it has been employed by various experimentalists, cf.~\cite{StruikPhD1977,Hastie1991,Pierik2020,Hodge1995,Hermida2018}.

\subsection{Physical Aging}\label{sec:PA}
Physical aging explains, why in the long-term creep of polymers, the observed creep compliance gradually deviates from the corresponding TTSP master curve. In his extensive studies on a number of synthetic polymers, Struik~\cite{StruikPhD1977} showed that the aging time $t_{\mathrm{e}}$, which elapses between quenching the polymer below its glass transition temperature and material testing, has a significant effect on the mechanical properties. This phenomenon is visible in the creep behavior of amorphous and semi-crystalline polymers and can be explained by the free volume theory of molecular motion. The theory roughly says that the molecular mobility of amorphous polymers at temperatures above glass transition, $T> T_{\mathrm{g}}$, is sufficient to immediately reach thermodynamic equilibrium, thanks to enough free volume. For $T< T_{\mathrm{g}}$, the molecular mobility becomes so small that there is a difference between the actual and the equilibrium free volume. This difference is larger the faster the polymer is cooled and acts as a driving force to reach the equilibrium state with the remaining molecular mobility. The required molecular rearrangements are slow, self-delaying, lead to time-dependent changes in mechanical properties, and come to a standstill at very low temperatures.

\begin{figure}[tb]
  \begin{minipage}[t]{0.49\textwidth}
    \centering
    \includegraphics[scale=1]{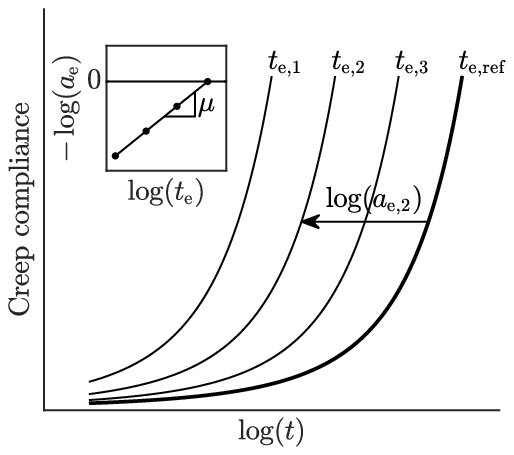}
    \caption{Isothermal creep curves at different aging times $t_{\mathrm{e}}$; the inset shows the curve shifts on the logarithmic time scale with respect to the reference curve with aging time $t_{\mathrm{e,ref}}$}
    \label{fig:LogaLogte}
  \end{minipage}\hfill
  \begin{minipage}[t]{0.49\textwidth}
    \centering
    \includegraphics[scale=1]{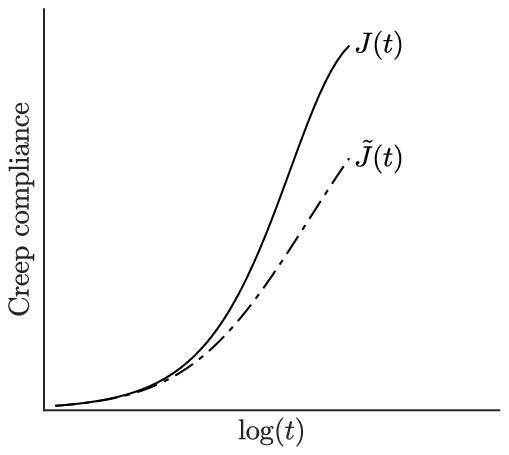}
    \caption{Prediction of long-term creep curve $\tilde{J}(t)$ obtained by correction of the momentary master curve $J(t)$ by means of the effective time theory}
    \label{fig:ETT}
  \end{minipage}
\end{figure}

Since molecular mobility depends directly on the available free volume, it decreases with aging time. This in turn increases the relaxation times, which is mapped by a certain shift factor $a_{\mathrm{e}}$. As a consequence, isothermal creep curves obtained at increasing aging times $t_{\mathrm{e}}$ are shifted to the right on the logarithmic time scale, Fig.~\ref{fig:LogaLogte}. This inverse proportionality between molecular mobility and aging time was confirmed by experimental results of isothermal short-term creep tests at small strains in \cite{StruikPhD1977}. On double-logarithmic representations of the shift factors $a_{\mathrm{e}}$ over the corresponding aging times $t_{\mathrm{e}}$, a linear relation was found, see inset of Fig.~\ref{fig:LogaLogte}. The slope of this relationship is denoted as the aging shift rate
\begin{align}\label{eqn:ShiftRate}
   \mu = -\frac{\Delta\log(a_{\mathrm{e}})}{\Delta\log(t_{\mathrm{e}})} \,.
\end{align}
With the sign convention according to literature, $-\log(a_{\mathrm{e}})$ denotes a horizontal curve shift to the right with respect to the reference curve. The shift rate is constant for a given temperature and has values of about one for a wide range of polymers.

To incorporate the physical aging phenomena in the macroscopic creep function \eqref{eqn:MaxwellCreepCurve} or its simplified form \eqref{eqn:KWW}, an effective time approach according to Struik~\cite{StruikPhD1977} will be utilized. In the effective time theory, the shift factor $a_{\mathrm{e}}$ is used to define a quasi-time function $\lambda(t)$ with
\begin{align}\label{eqn:dLambda}
  \td \lambda = a_{\mathrm{e}}(t,t_{\mathrm{e}}) \td t \,.
\end{align}
To integrate \eqref{eqn:dLambda}, the shift factor for a total aging time of $t_{\mathrm{e}}+t$ with respect to a reference elapsed aging time $t_{\mathrm{e}}$ needs to be specified. Here the aging shift rate \eqref{eqn:ShiftRate} is used to deduce
\begin{equation}
  a_{\mathrm{e}}(t,t_{\mathrm{e}}) = \left( \frac{t_{\mathrm{e}}}{t_{\mathrm{e}}+t} \right)^\mu \,.
\end{equation}
This model implies that all relaxation processes in the interval $\td t$ are slower by a factor of $1/a_{\mathrm{e}}$ at time $t>0$ than at time $t=0$. Thus, integration of \eqref{eqn:dLambda} gives
\begin{equation}\label{eqn:Lambda}
    \lambda(t) = \int_0^t a_{\mathrm{e}}(\xi,t_{\mathrm{e}}) \td\xi \,\stackrel{\mu\neq 1}{=}\, \frac{t_{\mathrm{e}}}{1-\mu}\left[\left(1+\frac{t}{t_{\mathrm{e}}}\right)^{1-\mu}-1\right] \,.
\end{equation}
The effective time function $\lambda(t)$ allows the conversion from the \textit{momentary} master curve $J(t)$ to the corrected long-term creep compliance $\tilde{J}(t)$ by the relation
\begin{equation}
  \tilde{J}(t) = J(\lambda) \,,
\end{equation}
see Fig.~\ref{fig:ETT}. When no aging takes places, i.e.~$\mu=0$, Eq.~\eqref{eqn:Lambda} corresponds to the momentary master curve creep time $t$.
 
To conclude this very brief introduction on the basic aspects of physical aging, it should be noted that these were initially developed for fully amorphus polymers. Here, the shift rate must be in the range $0.7<\mu<1$ to obtain reliable creep predictions over very long times \cite{StruikPhD1977}. However, Struik~\cite{StruikPhD1977,Struik1987-1} found that semi-crystalline polymers and filled rubbers are also affected by physical aging mechanisms at temperatures below the glass transition temperature of their amorphous phase, and the proposed effective time approach can be used  as well.

\section{Material and Methods}\label{sec:MandM}
In what follows, an overview of the specimen fabrication and their geometry are given. Then, the loading regimes of the three-point bending tests conducted are explained in detail.

\subsection{Test specimens}\label{sec:TestSpec}
\begin{figure}[b]
  \centering
  \includegraphics[scale=0.8]{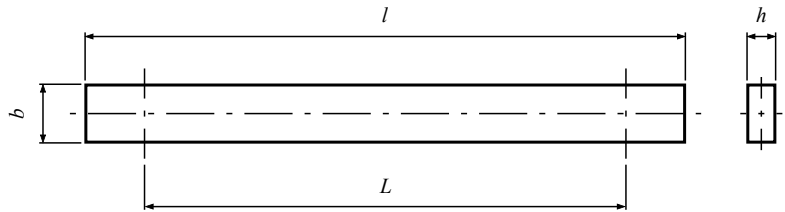}
  \caption{Specimen dimensions according to ISO 178 with nominal length $l=100\,$mm, width \mbox{$b=10$\,}mm, height $h=5\,$mm and support span $L=80\,$mm}
  \label{fig:SpecGeom}
\end{figure}

The geometry of the test specimens used in the flexural creep tests follows the ISO 178 standard and is shown in Fig.~\ref{fig:SpecGeom}. The specimens were printed in a semi-professional desktop 3D printer using PLA filament with the brand name Ultrafuse PLA from BASF 3D Printing Solutions and a diameter of 1.75\,mm. The used print settings are summarized in Tab.~\ref{Tab:PrintSettings}.

To improve adhesion between the printing platform and the printed object, it is recommended in practice
to set the platform temperature in the range of the glass transition temperature of the material. This increases the temperature in the first few layers of the object. In the layers atop, the effect of this heat source is not present, so the filament deposited here cools very quickly to the prevailing ambient temperature. For example, by means of a simplified heat transfer simulation of the cooling behavior of a deposited elliptical ABS fiber in a 3D printing process, Rodriguez~\etal~\cite{Rodriguez2000} found that it takes less than one second for its core temperature to fall below the glass transition temperature of 94\textcelsius, assuming two-sided contact (e.g.~left and bottom) to already printed material with a temperature of 55\textcelsius{} and a nozzle temperature of 270\textcelsius. Since the test specimens used in our study are low in height and consist of only 25 layers, their cooling behavior during printing would be significantly affected by elevated platform temperatures. To minimize this effect and to achieve high and uniform cooling rates, the temperature of the platform was set to a low value of 30\textcelsius, which is well below the glass transition temperature $T_{\mathrm{g}}=61$\textcelsius\footnote{Ultrafuse PLA, Technical Data Sheet v4.4, BASF 3D Printing Solutions GmbH} of the PLA material used. The bond between the platform and the printed sample was improved by using a water-soluble adhesive.

\begin{table}[tb]
  \setlength{\tabcolsep}{12pt}
  \centering
  \begin{tabular}{ll}
    \hline\noalign{\smallskip}
    print setting & value\\
    \noalign{\smallskip}\svhline\noalign{\smallskip}
    nozzle size & 0.4\,mm\\
    extruder temperature & 215\textcelsius\\
    platform temperature & 30\textcelsius\\
    layer height & 0.2\,mm\\
    print speed & 60\,mm/s\\
    shell count & 1\\
    infill pattern & 45\textdegree-90\textdegree~line pattern\\
    infill density & 100\,\%\\
    extrusion multiplier & 102\,\%\\
    cooling fan & always on\\
    \noalign{\smallskip}\hline\noalign{\smallskip}
  \end{tabular}
  \caption{3D print settings used for sample fabrication}
  \label{Tab:PrintSettings}
\end{table}

After a specimen was finished printing, it was left in the printer for about 10 minutes until the printing platform reached the ambient temperature of about 20\textcelsius. The printer was located in the same room where the creep tests were performed.

\subsection{Sequential creep tests}\label{sec:SeqCreepTests}
\begin{figure}[tb]
  \sidecaption
  \includegraphics[scale=1]{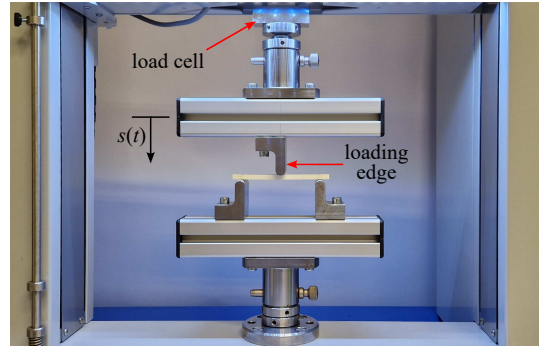}
  \caption{Three-point bending apparatus connected to a universal testing machine}
  \label{fig:TestSetup}
\end{figure}

To study the creep behavior of the test specimens, three-point flexural creep tests were conducted in accordance with ISO~899-2 and under the conditions presented in the following. In our experiments, we use a three-point bending apparatus connected to a universal testing machine as shown in Fig.~\ref{fig:TestSetup}. Prior to testing, the specimens were measured with a caliper gauge in width and height with an accuracy of 0.05\,mm.

\begin{figure}[b]
  \centering
  \includegraphics[scale=1]{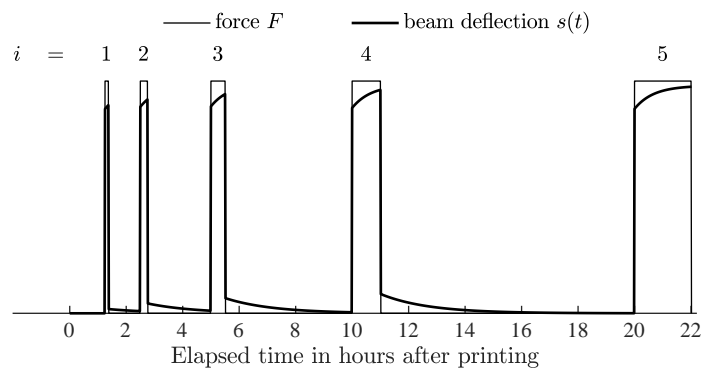}
  \caption{Sequential creep test procedure}
  \label{fig:SeqCreepTest}
\end{figure}

In the sequential creep tests, the specimens were subjected to a constant load $F$ for several short periods of time at aging times of $1.25,2.5,5,10$ and $20$ hours, see Fig.~\ref{fig:SeqCreepTest}. This sequential procedure was also used by Struik~\cite{StruikPhD1977} and others, where the duration $T_i$ of the $i$-th loading sequence is small compared to the previous aging time $t_{\mathrm{e},i}$. Thus, almost no additional aging occurs during the loading sequences and momentary creep curves are obtained. In our study, the durations $T_i$ were 10\,\% of the corresponding aging times $t_{\mathrm{e},i}$.

In order to analyze the individual creep sequences unambiguously in terms of the elapsed aging time, it must be ensured that the material properties do not change significantly as a result of the sequential loading. Therefore, the tests are carried out at small strains, where the assumption of linear viscoelasticity and Boltzmann's superposition principle are valid. In our tests, the specimens were loaded with a nominal force of 10\,N to obtain small flexural strains below 1\,\%. The force was measured with a load cell and kept nearly constant in a narrow range around the nominal value. The tests were performed in absence of any UV radiation at a constant temperature of about 20\textcelsius{} and relative air humidity between 30 and 50\,\%, which was confirmed by measurements during the tests.

Using this test setup in our experiments, the flexural creep strain $\varepsilon(t)$ of a loaded specimen was determined by the relation
\begin{equation}\label{eqn:epsilon}
  \varepsilon(t) = \frac{6h}{L^2}(s(t)-s_0) \,.
\end{equation}
Here $L$ denotes the span between the specimen supports, which must be a multiple of $h$ for Euler-Bernoulli beam theory to be applicable; $s(t)$ is the crosshead travel distance and $s_0<s(t=0)$ its absolute value at which the loading edge (see Fig.~\ref{fig:TestSetup}) has initial contact to the specimen. In practice, the value of $s_0$ was determined at the time when the measured force signal begins to differ from zero.

As a measure of the creep behavior of the specimens considered, their flexural creep compliance $J(t)$ was calculated as a function of time by Eq.~\eqref{eqn:DefCompliance}, where $\varepsilon(t)$ is the strain determined by \eqref{eqn:epsilon} and $\sigma(t)$ is the nearly constant flexural stress obtained by
\begin{equation}
  \sigma(t) = \frac{3 L}{2bh^2}F(t) \,.
\end{equation}  
Both strain and stress were calculated using the individual measured specimen dimensions $h$ and $b$, which differed slightly from the nominal values in Fig.~\ref{fig:SpecGeom}. The time at which a specimen is fully loaded is declared as $t_i=0$ and defines the initial creep compliance, $J_{i,0} = J_i(t_i=0)$, of the $i$-th creep sequence.

\subsection{Long term creep test}\label{sec:LtCreepTests}
In addition to the sequential creep tests, a long-term creep experiment was conducted over a loading period of one week. Here the specimen was subjected to a nominal force of 10\,N and held constant over 170\,h. The test was performed to compare its result in the short-term range with the short-term creep of specimens of the same age that had already been subjected to several loading sequences. Therefore, the long-term test was started after an elapsed aging time of five hours, which corresponds to the specimen age at the beginning of the third sequence in the sequential tests according to Sect.~\ref{sec:SeqCreepTests}.

\section{Experimental Results}\label{sec:Results}
In the following, the steps of post-processing of the experimental data obtained with the methods described in \ref{sec:SeqCreepTests}--\ref{sec:LtCreepTests} are detailed. A qualitative discussion of the final results presented can be found in Sect.~\ref{sec:Disc}.

\subsection{Sequential creep tests}\label{sec:SeqCreepTestResults}
\begin{figure}[b]
  \centering
  \includegraphics[scale=1]{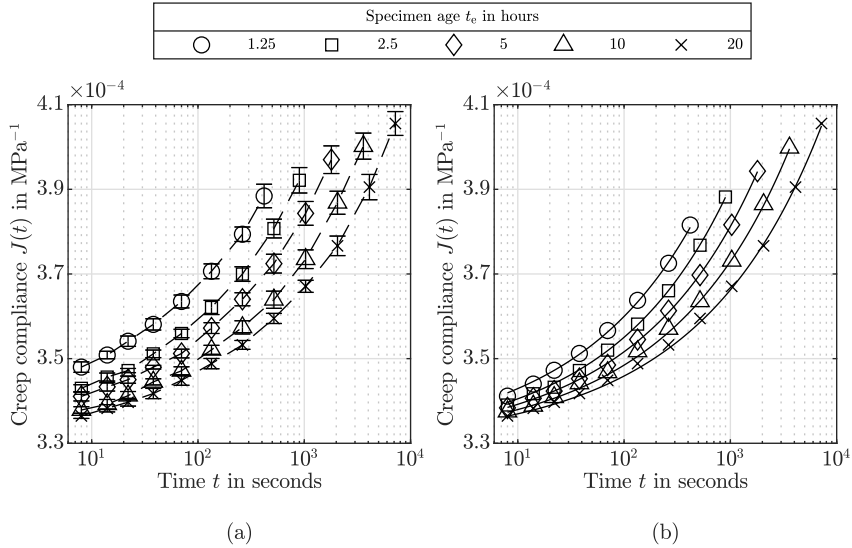}
  \caption{\textbf{a}) Calibrated creep curves averaged over five specimens; \textbf{b}) creep curves shifted vertically to the reference initial compliance $\bar{J}_{5,0}$ and fitted individually to the creep model \eqref{eqn:KWW}}
  \label{fig:CurveComb}
\end{figure}

In total five specimens were printed and tested according to the procedures explained in sections \ref{sec:TestSpec}--\ref{sec:SeqCreepTests}. Fig.~\ref{fig:CurveComb}a shows the averaged creep compliance curves obtained by sequential creep tests for different aging times. The means of the initial creep compliances, $\bar{J}_{i,0}$, were computed for each set, $i=1,...,5\,$, and all curves were calibrated accordingly. The error bars in Fig.~\ref{fig:CurveComb}a refer to the remaining scatter of the calibrated curves, given as standard deviation.

Plotted in a semi-logarithmic diagram, the averaged creep curves show a similar shape and a nearly equidistant distribution with respect to the logarithmic time scale. The higher the specimen age, the more they are shifted to the right, which can be explained by an increase in relaxation times. In view of the explanations in Sect.~\ref{sec:PA}, it is therefore plausible to assume physical aging as the cause of the observed creep behavior.

The 20\,h aged curve is chosen as reference curve, $t_{\mathrm{e,ref}}=t_{\mathrm{e},5}=20$\,h. To quantify the curve shifts on the logarithmic time scale as a function of aging time $t_{\mathrm{e}}$, the averaged creep curves were first arranged by vertical shifts $\Delta \bar{J}_i$ so that their initial compliances $\bar{J}_{i,0}$ coincide with the initial compliance $\bar{J}_{5,0}$ of the reference curve. Next, the experimental data calibrated this way, represented by symbols in Fig.~\ref{fig:CurveComb}b, were individually fitted to the creep model~\eqref{eqn:KWW} using least squares minimization with fitting parameters $\tau$ and $m$. The resulting fit curves are shown as continuous lines in Fig.~\ref{fig:CurveComb}b. The corresponding fits of parameter $m$ are listed in Tab.~\ref{Tab:FitTab}. Since these values differ only slightly from each other, it can be verified that all creep curves are characterized by the same shape and differ only by retardation time $\tau_i$. Thus, parameter $m$ is assumed to be a material constant and set to the mean value of $m=0.35$ for subsequent analysis.

\begin{table}[tb]
  \setlength{\tabcolsep}{6pt}
  \centering
  \begin{tabular}{lclccccc}
    \hline\noalign{\smallskip}
    creep sequence & $i$ & 1 & 2 & 3 & 4 & 5 & unit\\
    \noalign{\smallskip}\svhline\noalign{\smallskip}
    specimen age & $t_{\mathrm{e},i}$ & 1.25 & 2.5 & 5 & 10 & 20 & h\\
    vertical shift to $\bar{J}_{5,0}$ & $\Delta \bar{J}_i$ & -0.69 & -0.39 & -0.27 & -0.04 & 0 & $10^{-5}$\,MPa\textsuperscript{-1}\\
    model parameter (1st fit) & $m_i$ & 0.35 & 0.36 & 0.35 & 0.35 & 0.34 & -\\
    retardation time (2nd fit) & $\tau_i$ & 1.09 & 1.40 & 2.38 & 4.19 & 7.58 & $10^{5}$\,s\\
    \noalign{\smallskip}\hline\noalign{\smallskip}
  \end{tabular}
  \caption{Shift and fit results from data analysis}
  \label{Tab:FitTab}
\end{table}

Given a constant value for $m$, the shift $\log(a_i)$ of a creep curve $J_i(t)$ relative to a reference curve $J_{\mathrm{ref}}(t)$, each represented by \eqref{eqn:KWW}, can be calculated by the relation
\begin{equation}\label{eqn:logae}
  -\log(a_i) = \log\left(\frac{\tau_i}{\tau_{\mathrm{ref}}}\right)\,.
\end{equation}
To calculate the shifts $\log(a_i)$ of the considered creep curves with respect to the chosen reference curve $\bar{J}_5(t)$, they were again fitted to the creep model \eqref{eqn:KWW} with fitting parameters $\tau_i$ and the now constant model parameter $m=0.35$. The fitted values of $\tau_i, i=1,...,5$ are listed in Tab.~\ref{Tab:FitTab}; they were used to calculate $\log(a_i), i=1,...,5$ by Eq.~\eqref{eqn:logae}.

\begin{figure}[b]
  \begin{minipage}[t]{0.49\textwidth}
  \centering
    \includegraphics[scale=1]{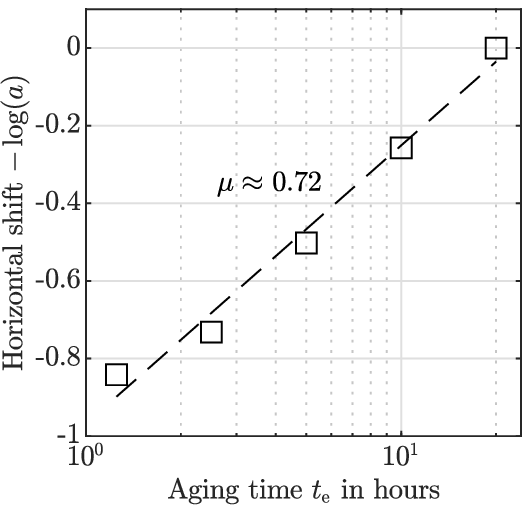}
    \caption{Calculated time shifts as a function\\ of specimen age $t_{\mathrm{e}}$}
    \label{fig:Shifts}
  \end{minipage}\hfill
  \begin{minipage}[t]{0.49\textwidth}
  \centering
    \includegraphics[scale=1]{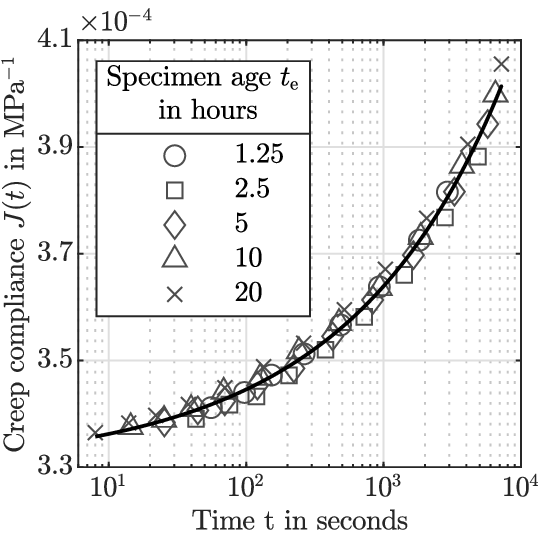}
    \caption{Experimental creep curves shifted to the reference aging time $t_{\mathrm{e,ref}}=20$~h}
    \label{fig:Superpos}
  \end{minipage}
\end{figure}

The results are shown in Fig.~\ref{fig:Shifts} over the aging time $t_e$ in a double-logarithmic plot. The data points follow, to a good approximation, a linear relationship with slope $\mu=0.72$ representing the shift rate~\eqref{eqn:ShiftRate} calculated by linear regression. Fig.~\ref{fig:Superpos} shows the experimental creep curves each correspondingly shifted by $\log(a_i), i=1,...,5$ to the reference aging time $t_{\mathrm{e,ref}}=20$\,h. The solid line represents the fit result for the reference creep curve~$\bar{J}_5(t)$.

\subsection{Long term creep test}\label{sec:LtCreepTestResults}
\begin{figure}[h]
  \centering
  \includegraphics[scale=1]{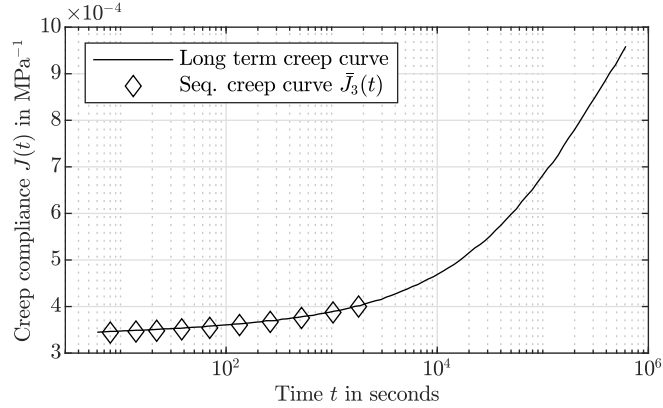}
  \caption{Comparison of the averaged short-term creep curve $\bar{J}_3(t)$ with a long-term creep curve obtained using a single specimen with initial aging time $t_{\mathrm{e}}=5$\,h.}
  \label{fig:CompCurves}
\end{figure}

The creep curve obtained from the long-term experiment is shown as a solid line in Fig.~\ref{fig:CompCurves}. The diamond-shaped data points refer to the averaged short-term creep curve $\bar{J}_3(t)$ calculated as described in Sect.~\ref{sec:SeqCreepTestResults}. To compare both curves with respect to their position on the logarithmic time scale, the short-term curve was shifted slightly vertically so that their initial compliances coincide. It can be clearly seen that the curves are not shifted against each other on the logarithmic time scale and have an identical shape.

The long-term creep curve shows that even after a one-week load, a state of equilibrium has not yet been reached and the material continues to creep. At the end of the recorded period, a creep modulus of 1040\,MPa is calculated, which corresponds to only about 56\,\% of the quasi-static flexural elastic modulus of 1860\,MPa given in the material data sheet.

\section{Discussion}\label{sec:Disc}
Sequential creep tests were performed to analyze the short-term flexural creep behavior of 3D printed, increasingly aged specimens, the results of which are presented in Sect.~\ref{sec:Results}. Plotted on a logarithmic time scale, the obtained short-term creep curves show similar shapes but are horizontally shifted as a function of aging time. Their relative shifts to a reference curve were found to satisfy equation~\eqref{eqn:ShiftRate} with a calculated shift rate $\mu\approx 0.72$ to a good approximation. 

By comparing two creep curves obtained on identically aged specimens with different preloading histories, see Sect.~\ref{sec:LtCreepTestResults}, it was shown that sequential loading of specimens with small forces as described in Sect.~\ref{sec:SeqCreepTests} does not significantly affect their creep behavior. Since all creep tests were performed at constant temperatures and under the exclusion of UV radiation, we assume that physical aging is the main reason for the observed increase in retardation times as a function of specimen age. Post-crystallization was also excluded as a cause since such crystallization is only to be expected for semi-crystalline polymers that are exposed to temperatures above their glass transition temperature for extended time periods \cite{Kalinka1995}.

Struik~\cite{StruikPhD1977} argued and showed experimentally that physical aging occurs in a certain temperature range below the glass transition temperature and disappears at both very low and high temperatures. Here the shift rate tends to zero and reaches a maximum at a temperature between these boundary states. From the test results obtained at an ambient temperature of 20\textcelsius{} we calculated a shift rate of $\mu\approx 0.72$, which is far from zero, but not very close to one. Since the testing temperature was about 40\textcelsius{} lower than the glass transition temperature of the PLA material used, we thus assume that the maximum possible shift rate is larger than the value we calculated and is reached at higher temperatures.

Motivated by Struik's argumentation that physical aging is not limited to individual polymers but is a more general phenomenon, we further assume that other 3D printed thermoplastics are also affected by the same physical aging shown here using PLA as an example material. 
The strength of this aging influence depends on the glass transition temperature of the material printed, the ambient temperature during the creep test, and the cooling rate achieved during extrusion. It should be noted that the cooling process in 3D printing differs significantly from that in conventional extrusion processes, such as injection molding. Whereas in the latter, inhomogeneous temperature gradients can lead to locally different cooling conditions, these can be assumed to be approximately identical over the entire component in 3D printing. The creep and aging properties determined on small 3D printed test specimens can therefore be directly transferred to large 3D printed components under otherwise identical conditions.

Since physical aging is temperature-dependent, special temperature treatments in advanced 3D printing, e.g. the use of heated print chambers or subsequent annealing, c.f.~\cite{Akhoundi2020}, would have a significant influence on the long-term creep behavior of printed parts. In our prints, the cooling rate was just large enough to achieve a shift rate $\mu > 0.7$, which according to \cite{StruikPhD1977} allows valid prediction of aging-influenced long-term creep behavior based on momentary TTSP master creep curves as described in Sect.~\ref{sec:PA}. However, further research is needed to investigate in detail the extent to which advanced heat treatments affect physical aging and thus the long-term creep of 3D printed thermoplastics.

To demonstrate that 3D printed thermoplastics creep significantly over long time periods even at temperatures well below their glass transition temperature, a long-term creep experiment was conducted over a week at a constant temperature. The corresponding results are presented in Sect.~\ref{sec:LtCreepTestResults} and are essential to consider when designing load-bearing structures made of 3D printed thermoplastics. We refer in particular to those applications where precise knowledge of the viscoelastic material properties is important, e.g. in the design of interference fits for shafts or ball bearings. 

Using the test setup described in Sect.~\ref{sec:SeqCreepTests}, creep curves were obtained by calculating the creep compliance as a function of the load the specimens were subjected to and the time-dependent change in the crosshead travel distance of the testing machine. Therefore, no additional strain gauges or extensometers were used. However, the creep curves 
obtained by this method exhibited considerable scatter in the calculated initial compliances, even when recorded at identical aging times. Thus, in order to use them for a quantitative interpretation of the specimens' creep behavior, some corrections had to be made as described in Sect.~\ref{sec:SeqCreepTestResults}. The reasons for the observed scatter in the initial compliances are, in our opinion, related to the measurement of the reference crosshead travel distance $s_0$, see Eq.~\eqref{eqn:epsilon}, which is affected by changing contact conditions between the loading edge and the specimen and by possibly inaccurate force measurements in the range of very small loads. For an accurate determination of the initial compliances of sequentially loaded creep specimens, our proposed test setup can therefore only be recommended to a limited extent. To improve the measurement results and to reduce the post-processing effort, we recommend the use of high-resolution load cells for the accurate measurement of small forces or direct strain measurements with strain gauges or extensometers, cf. \cite{Hastie1991,Pierik2020}.

\bibliographystyle{spmpsci}
\bibliography{references}

\end{document}